\documentclass{KapProc}
\usepackage{graphics}

\let\footnote\savefootnote
\let\footnotetext\savefootnotetext 
\let\footnoterule\savefootnoterule 

\kluwerbib

\begin{document}

\articletitle[]{The very lithium rich post-AGB SB2 binary HD\,172481}
\noindent {\small\em To appear in the proceedings of the conference:
``Post-AGB Objects (Proto-Planetary Nebulae) as a Phase of Stellar Evolution'',
held in Toru\'n, Poland, July 5-7, 2000, eds. R. Szczerba, R. Tylenda, and S.K.
Gorny.}
\chaptitlerunninghead{HD\,172481}
\author{Maarten Reyniers, Hans Van Winckel}
\affil{Instituut voor Sterrenkunde, Celestijnenlaan 200B, 3001 Heverlee, Belgium}
\email{Maarten.Reyniers@ster.kuleuven.ac.be}

\begin{abstract}
Double lined spectroscopic binaries in an evolved stage of evolution are
expected to be extremely rare since they must consist of equally luminous
and thus almost equally evolved objects, which requires an extremely similar
initial mass. In this contribution we discuss such rare double evolved SB2
system: HD\,172481. This binary includes an F-type post-AGB object and
an M-type AGB companion. The spectrum shows a surprisingly strong
Li\,I 670.8\,nm line with an equivalent width of $W_{\lambda}$\,=\,54\,m\AA,
yielding a lithium abundance of $\log\epsilon({\rm Li}) = 3.6$. Several
explanations for this huge lithium content are explored.
\end{abstract}


\section{HD\,172481 in the post-AGB sample}
In the course of our ongoing program to study the chemical composition of
post-AGB stars (see Van Winckel, this volume), high resolution, high
signal-to-noise spectra were taken of the F2Ia supergiant HD\,172481.
This star was previously selected as a candidate post-AGB star by Oudmaijer et
al. (1992) because of its far infrared excess and its high galactic latitude
($b$\,=\,$-10.37^{\circ}$).

\section{Observational data}\label{sect:obdata}
Our {\bf photometric} data consist of 54 data points in the Geneva photometric
system between 1989 and 1996, near-IR JHKLM photometry taken with the ESO\,1m
telescope in 1992 and the IRAS-fluxes.
62 {\bf radial velocity} measurements were obtained with the CORAVEL radial
velocity spectrometer (Danish\,1.5m telescope, La Silla, Chile) between 1983
and 1997 and 4 recent measurements with the CORALIE spectrograph (Euler
telescope, La Silla, Chile).
5 {\bf high resolution, high signal-to-noise spectra} were taken with several
spectrographs and telescopes: 2 spectra with the UES (WHT, La Palma,
Spain) in 1995, 2 with the EMMI spectrograph (NTT, La Silla, Chile) in 1997
and 1998 and 1 with FEROS (ESO\,1.5m, La Silla, Chile) in 2000.

\section{HD172481 as a binary}
\begin{figure}[t]\caption{The 1998\,EMMI (top) and FEROS (bottom) spectra in the 700-715\,nm TiO band head region. The top spectrum is offset by 0.2 units. The TiO bands in the red spectrum of HD\,172481 are a clear signature of its cool companion. Moreover, these bands vary in time as evidenced by our 1998 EMMI spectrum. Band head wavelengths are from Valenti et al. (1998).}\label{fig:TiO}
\resizebox{\hsize}{!}{\rotatebox{-90}{\includegraphics{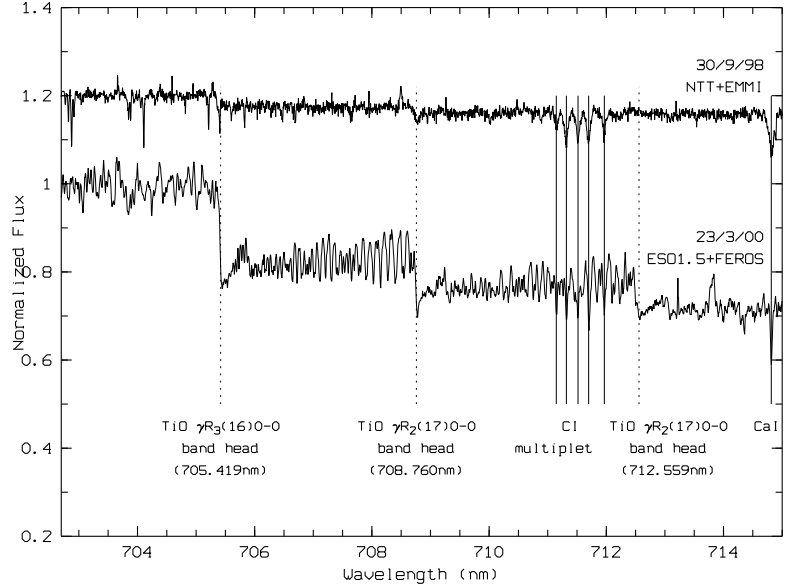}}}
\end{figure}

A comparison of the blue and green regions of the spectra of this object with
spectra of the normal massive F0Ib supergiant HR\,1865 confirms the F2Ia0
spectral type given by the SIMBAD database. However, our 1997\,EMMI and FEROS
spectra show clear and strong discontinuous jumps at certain wavelengths in the
red region which we identified as {\bf TiO band heads} (Fig. \ref{fig:TiO}).
Since the pulsational amplitude is very low ($\Delta$V\,$\sim$\,0.2), we
conclude that these bands are caused by a cool M-type companion. The red region
of the spectrum of HD\,172481 is therefore a mixture of two spectral types:
atomic lines induced by a hot component ($\sim$7250\,K) and TiO bands caused
by a cool component ($\sim$3500\,K). Moreover, the strength of these bands is
clearly variable in time (also Fig. \ref{fig:TiO}).

\begin{figure}
\begin{minipage}[t]{5.8 cm}
\caption{The dereddened SED of HD\,172481 with a E(B-V)\,=\,0.44. The dotted line represents the Kurucz model with T$_{\rm eff}$\,=\,7250\,K and log(g)\,=\,1.5; the dashed line the model with T$_{\rm eff}$\,=\,3500\,K and log(g)\,=\,0.5; the full line is the combined model.}\label{fig:SED}
\resizebox{\hsize}{!}{\includegraphics{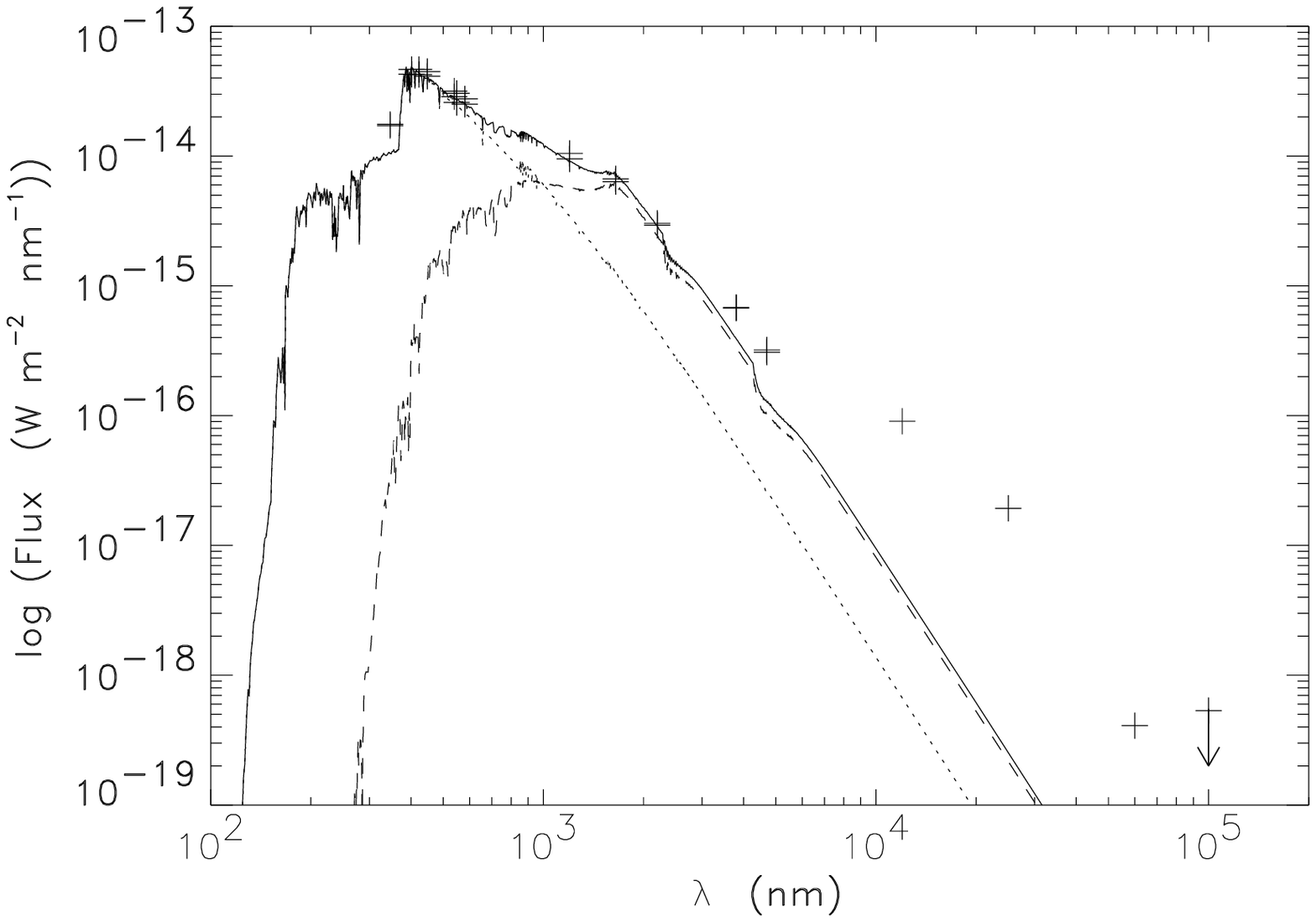}}
\end{minipage}
\hfill
\begin{minipage}[t]{5.8 cm}
\caption{The abundances of HD\,172481 relative to iron [el/Fe].
Errorbars are given if the number of lines is more than 3. The horizontal line
represents [el/Fe]=0. Solar abundances are mainly taken from Grevesse (1989).}\label{fig:elfe}
\resizebox{\hsize}{!}{\includegraphics{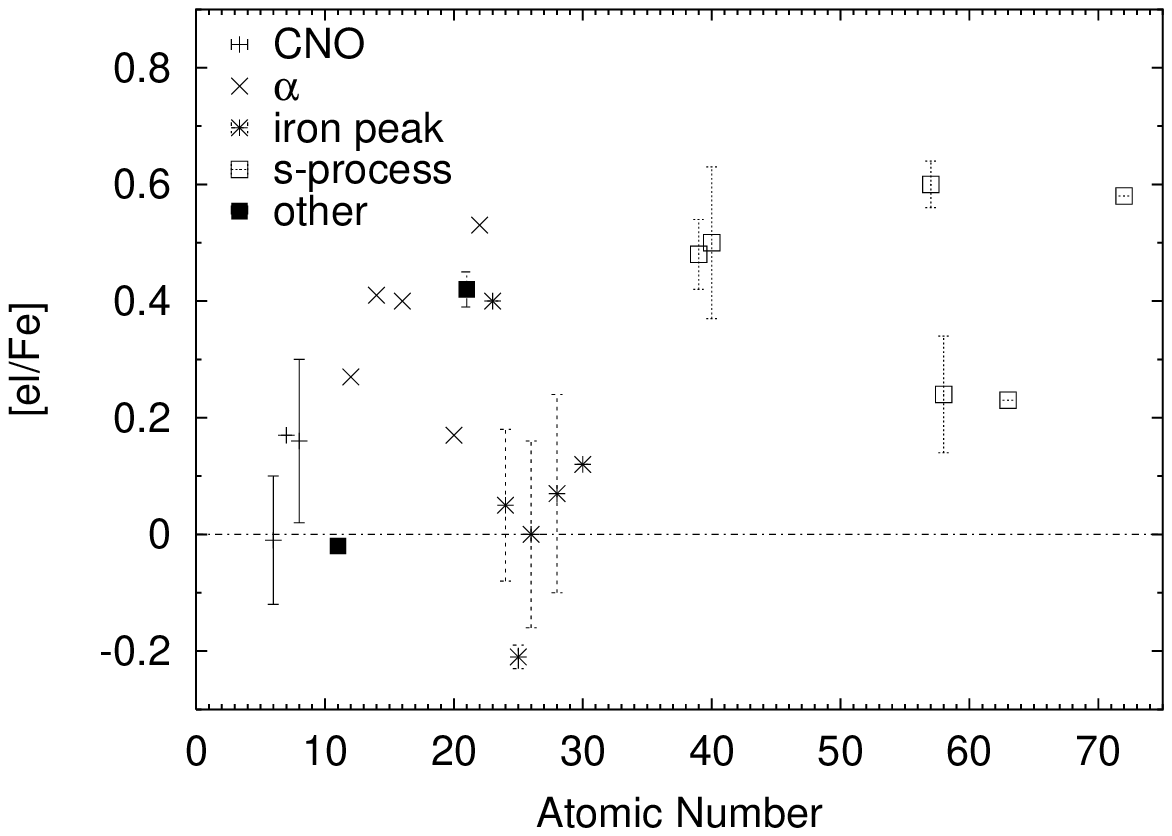}}
\end{minipage}
\end{figure}

Other evidence confirming the binary nature of HD\,172481 was found in its
{\bf spectral energy distribution} (SED hereafter), which is constructed from
the collected photometry in Fig. \ref{fig:SED}
with an adopted reddening of E(B-V)=0.44. The fit of only the F-type (Kurucz)
model atmosphere clearly did not satisfy from about $\lambda$\,$\sim$10$^3$\,nm
redwards. Therefore, we had to invoke a second, cool component with
T$_{\rm eff}$\,=\,3500\,K. The latter temperature is, however, highly uncertain
because it is derived by only a few photometric points.

Two more interesting remarks can be made concerning the SED. The
{\em luminosity ratio} of both components L$_{\rm F-type}$/L$_{\rm M-type}$ is
found to be 1.8. This ratio is rather dependent on the
adopted reddening: using E(B-V)\,=\,0.2, it equals to 1. In any case,
the two components must be of very similar luminosity. Consequently,
also the M-type component must be very luminous and probably in its AGB
phase of stellar evolution. A second remark concerns the IRAS-fluxes
(the 4 most redward points on the SED) pointing to a cool infrared excess
which we interpret as caused by a circumstellar dust shell or disk.

The binary nature of HD\,172481 could, however, not be resolved by our
{\bf radial velocity monitoring}: although a peak-to-peak amplitude of
17\,km\,s$^{-1}$ is observed, we could not discover a long-term trend in the
radial velocity data. A possible explanation for this fact can be an
unfavourable inclination angle, but most likely it is due to a long orbital
period. The variation induced by the orbital motion is then completely washed
out by the (low amplitude) pulsation. This (semi-regular) low-amplitude
pulsation in also seen in our Geneva photometry.

\section{Chemical Analysis}
A detailed chemical abundance analysis was carried out based on the high
resolution spectra. Atmospheric parameters were obtained with the commonly used
spectroscopic method: excitation equilibrium for the effective temperature
(T$_{\rm eff}$), ionisation equilibrium for the gravity ($\log(g)$) and the
microturbulent velocity ($\xi_t$) is obtained by demanding that the iron
abundance is independent of the reduced equivalent widths
($W_{\lambda}/\lambda$) of the Fe\,I-lines.
This method resulted in T$_{\rm eff}$=\,7250\,($\pm$250)\,K,
$\log(g)$\,=\,1.5\,($\pm$0.5) (cgs) and
$\xi_t$\,=\,4\,($\pm$1)\,km\,s$^{-1}$.
48 Fe\,I and 14 Fe\,II lines were involved in
this determination. We used the atmospheric models of Kurucz (CD-ROM set, 1993)
in combination with his abundance calculation program {\sc width9}. Our
linelist contains only lines with well determined oscillator strengths in detail
described in Van Winckel \& Reyniers (2000).

209 lines of 28 different ions were measured. We obtained a metallicity of
[Fe/H]\,=\,$-$0.55. The results of our analysis are graphically presented in
Fig. \ref{fig:elfe}. On this figure, one can clearly
see that\\
$-$\ the CNO elements are clearly {\em not} enhanced;\\
$-$\ the $\alpha$ elements are enhanced by $\sim$0.3 to 0.4\,dex, which is not
an intrinsic enhancement, but a galactic evolution effect for this metallicity
range;\\
$-$\ the other iron peak elements do follow the iron deficiency of
[Fe/H]\,=\,$-$0.55;\\
$-$\ the light s-process elements Y and Zr and the heavy s-process element La
are enhanced by $\sim$0.5 dex.

\section{Post-AGB status of HD\,172481}
In this section we collect our arguments for the proposed post-AGB status
of HD\,172481.\\
1. The high galactic latitude ($b$\,=\,$-10.37^{\circ}$),\\
2. the large radial velocities ($<$$v$$>$$\sim$\,$-$85\,km\,s$^{-1}$) and \\
3. the moderate metal deficiency ([Fe/H]=$-$0.55)\\
point to a population II membership. Furthermore,\\
4. the SED (Fig. \ref{fig:SED}) shows the presence of dust,
as a result of previous mass loss,  probably during the AGB phase.\\
5. The photometry and the H$\alpha$ profiles on our 4 spectra that include
H$\alpha$ show a variability very similar to other post-AGB stars ({\em cfr.}
L\`ebre, this volume).\\
6. Finally, the slight s-process overabundances are probably the result of
the third dredge-up, when He burning products are brought to the surface.
This last argument is, however, not straightforward. If we expect the signature
of a 3rd dredge-up, also the CNO elements should be enhanced, which is {\em not}
detected: the total CNO abundance is equal to the initial value.

\section{Lithium in HD\,172481}
\begin{figure}[t]\caption{The Li\,I resonance line in HD\,172481.
{\em left panel:} The UES and FEROS spectrum around 671\,nm with the Li\,I
doublet of 54\,m\AA. The top spectrum is offset by 0.3 units. The contribution
of the cool companion is larger for our FEROS run as evidenced by the detection
of the (weak) TiO band head at 671.447\,nm and the apparently lower s/n for this
spectrum.
{\em right panel:} Synthesis of the Li\,I resonance line using our UES spectrum.
A velocity broadening ($\zeta$) of 14\,km\,s$^{-1}$ is used in this synthesis,
deduced by the synthesis of unblended lines with a smooth profile.
The points are the observed spectrum, the lines  synthetic spectra with
$\log\epsilon({\rm Li})$\,=\,3.47, 3.57 and 3.67 resp. A least-squares fit
resulted in $\log\epsilon({\rm Li})$\,=\,3.57 (full line).}\label{fig:li}
\resizebox{11.5 cm}{7.5 cm}{\includegraphics{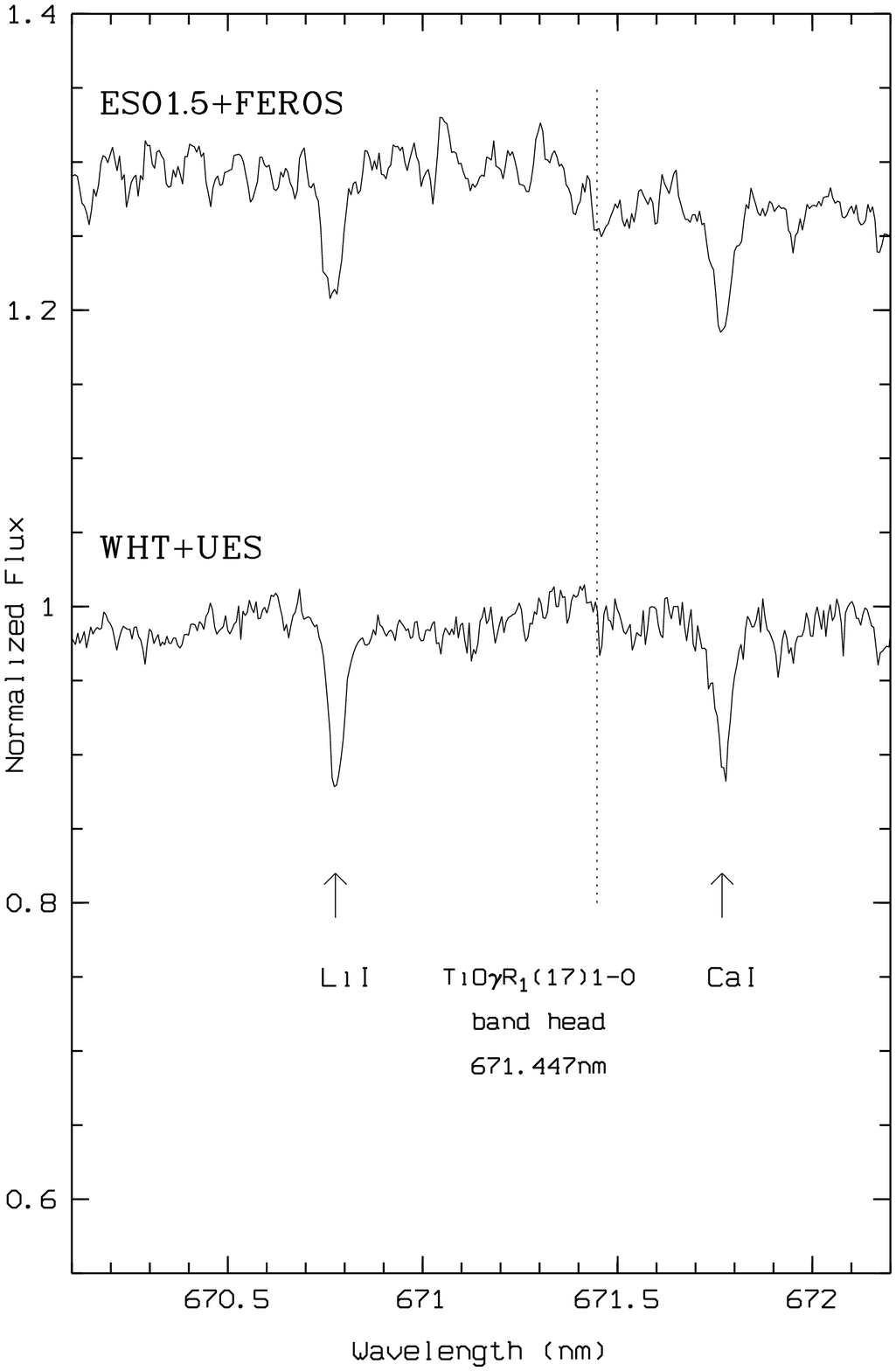}\includegraphics{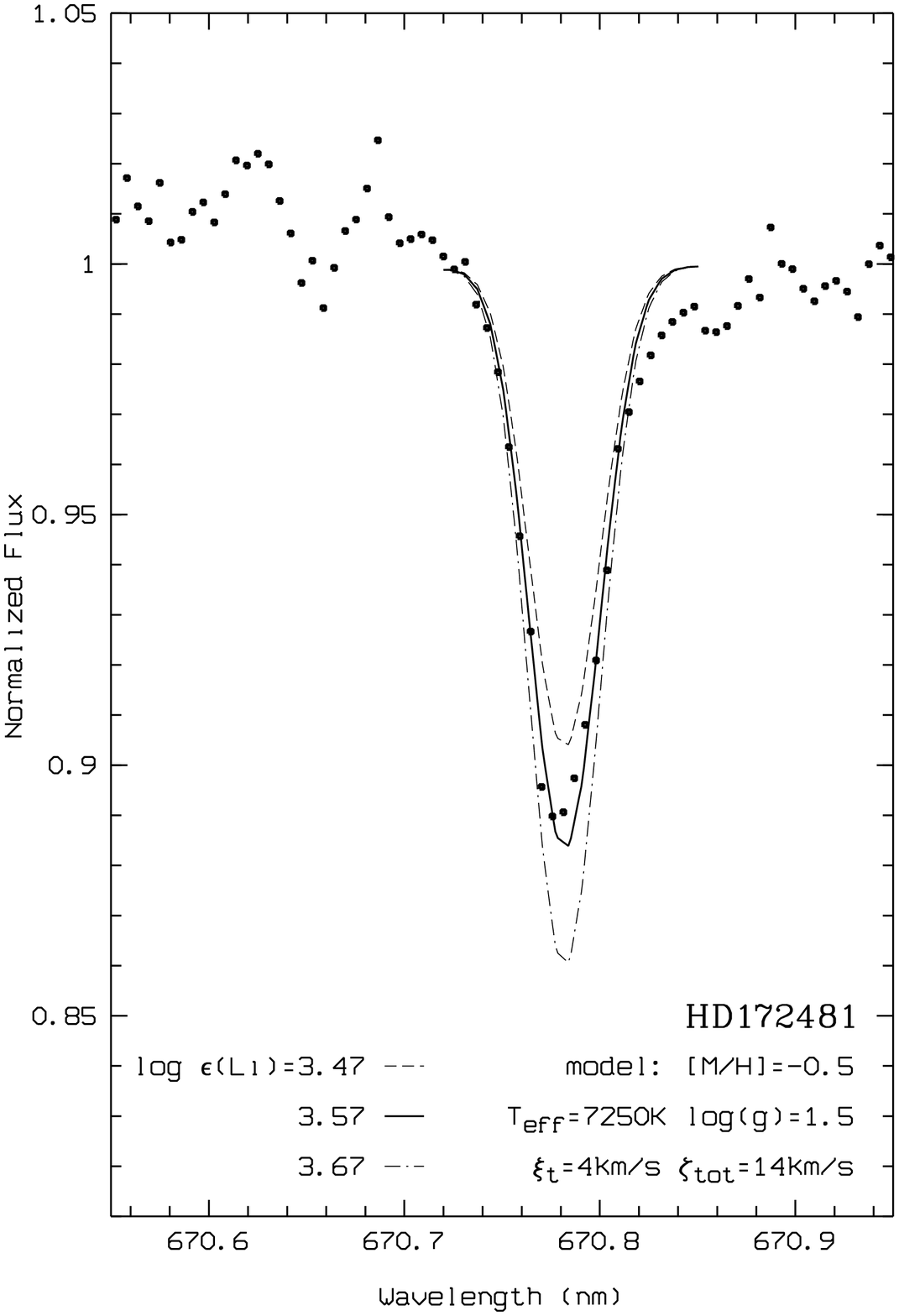}}
\end{figure}

The most surprising discovery during our spectral analysis was the detection of
a strong Li\,I resonance line at 670.8\,nm (Fig. \ref{fig:li}). Kurucz's
synthesis program {\sc synthe} was used to account for the doublet structure
of this line. A least-squares fit between the observed and the synthetic
spectrum yielded an abundance of $\log({\rm Li})$\,=\,3.57\,$\pm$\,0.2
(Fig. \ref{fig:li}). It is far from clear how HD\,172481 can be so Li rich,
exceeding even the interstellar medium value of
$\log\epsilon({\rm Li})$\,=\,3.3. During the evolution of a star, lithium is
destroyed, as it burns at relatively low temperatures. Lithium destruction and
depletion is expected to occur during the main-sequence evolution (e.g.
$\log({\rm Li})_{\odot}$\,=\,1.16) as well as during the RGB and AGB evolution,
as the deeper convective envelope reaches into lithium-depleted interior layers.

Nevertheless Li rich evolved stars exist. In general three classes of Li rich
evolved (AGB and post-AGB) stars can be distinguished: (1) the
luminous ``hot bottom burning'' (HBB) S-type AGB stars in the Clouds (Plez et
al. 1993, Smith et al. 1995) and the Galaxy (Garc\'{\i}a-Lario et al. 1999);
(2) galactic low-mass C stars (Abia et al. 1993; Abia \& Isern 1997, 2000)
which are mostly J-type and (3) 4 metal-poor low-mass post-3rd dredge-up objects
being IRAS\,22272+5435 (Za\v{c}s et al. 1995), IRAS\,05341+0852
(Reddy et al. 1997), IRAS\,Z02229+6208 and IRAS\,07430+1115 (Reddy
et al. 1999). Only for the HBB stars, Li production is supported by the models 
via the $^7$Be-transport mechanism (Cameron \& Fowler 1971). All models suggest
that HBB and the associated Cameron-Fowler mechanism only works in
intermediate mass stars (4-7\,M$_\odot$; e.g. Sackmann \& Boothroyd 1992).

For classes (2) and (3), which are most probably low-mass stars, no envelope
burning is predicted in any AGB model. By its old nature, and hence low mass,
a ``hot bottom burning'' scenario is not applicable to HD\,172481.
The most promising models to date, which could shed some light on the Li rich
evolved low-mass stars are the ``cool bottom processing'' models by Wasserburg
et al. (1995) and Sackmann \& Boothroyd (1999). These models are successfully
developed to explain the Li rich K giants (e.g. de La Reza et al. 1997 and
references therein) using a non-standard mixing mechanism. The same group of
authors suggests that their models might also work in the TP-AGB phase of
low-mass stars (Boothroyd \& Sackmann 1999).

\section{Conclusions}
In this contribution we shortly described our study on the very peculiar
supergiant HD\,172481. The TiO band heads superimposed on a typical F-type
spectrum reveal that it is a binary object, which is further strengthened by
the SED. The high radial velocities, the high galactic latitude and the moderate
metal deficiency found in our chemical analysis suggest that the F-type
component is an old and hence low-mass object. It is probably in its post-AGB
phase of evolution as evidenced by the dust, the nature of the (low amplitude)
variability and the slight s-process overabundances. Above all, this
star is found to be super rich in lithium, with a Li abundance of
$\log\epsilon({\rm Li}) = 3.6$. A more detailed description can be found in
Reyniers \& Van Winckel (2000).


\begin{acknowledgments}
\noindent Both authors acknowledge financial support of the Fund for Scientific Research - Flanders.
\end{acknowledgments}

\end{document}